\documentclass[preprint,12pt]{elsarticle}
\usepackage[utf8]{inputenc}
\usepackage{graphicx}
\usepackage{subfigure}

\usepackage{amssymb}
\usepackage{amsthm}
\usepackage{url}
\usepackage{amsmath}
\usepackage{subfigure}
\usepackage{textcomp}
\usepackage{xcolor}

\usepackage{amsmath,amssymb,graphicx,booktabs}
\usepackage{hyperref}
\hypersetup{colorlinks=true,linkcolor=blue,citecolor=blue,urlcolor=blue}

\journal{---}

\begin{document}

\begin{frontmatter}

\title{Directional Electrical Spiking, Bursting, and Information Propagation in Oyster Mycelium Recorded with a Star-Shaped Electrode Array}

\author[1]{Andrew Adamatzky}
\address[1]{Unconventional Computing Laboratory, University of the West of England, Bristol, UK}
\ead{andrew.adamatzky@uwe.ac.uk}

\begin{abstract}
Electrical activity in fungal mycelium has been reported in numerous species and experimental
contexts, yet its spatial organisation and propagation remain insufficiently characterised.
In this study we investigate the spatiotemporal structure of electrical potential dynamics in
oyster mushroom (\textit{Pleurotus ostreatus}) mycelium colonising a wood-shavings substrate.
Electrical signals were recorded using an eight-channel star-shaped differential electrode
array providing angular resolution around a central region of colonised substrate. We analyse
spike statistics, bursting behaviour, inter-channel correlations, and event-based propagation
delays. The results reveal strong directional heterogeneity in spiking frequency and amplitude,
clustered bursting dynamics, partial and localised coupling between channels, and reproducible
propagation patterns across spatial sectors. Electrical bursts originate preferentially in
specific directions and recruit other regions with 
with characteristic delays ranging from seconds to minutes to hours. These findings support the interpretation of fungal mycelium as a spatially
extended excitable medium capable of slow, distributed electrical signalling and signal
integration.
\end{abstract}


\end{frontmatter}

\section{Introduction}

Fluctuations in electrical potential and spike-like events have been observed in a wide range of
fungal species and substrates~\cite{slayman1976action,olsson1995action,adamatzky2018spiking,adamatzky2022language,dehshibi2021electrical,fukasawa2024electrical,buffi2025electrical,fukasawa2025electrical}. Such activity has been associated with physiological processes including growth, nutrient transport, environmental sensing, and responses to mechanical or chemical perturbations~\cite{gow1984transhyphal,harold1985fungi,gow1989relationship,gow1995electric,feng2019analysis,phillips2023electrical,fukasawa2023electrical}. While the existence of fungal electrical signalling is now well
established, most studies rely on recordings from single electrodes or linear electrode arrangements, limiting insight into the spatial organisation of electrical dynamics.

Understanding whether fungal electrical activity is spatially structured, directionally organised, or capable of propagation across a mycelial network is essential for  its functional significance. If electrical signals are generated locally and remain uncorrelated, they may reflect purely local physiological processes. Conversely, reproducible spatial correlations and propagation patterns would indicate distributed signalling and coordination across the network.

In this paper we address this question by recording electrical activity in oyster mushroom (\textit{Pleurotus ostreatus}) mycelium using a star-shaped electrode arrangement that provides angular resolution around a central region of colonised substrate. This geometry enables directional analysis of spiking activity, bursting behaviour, coupling, and propagation without
assuming a preferred linear axis. We report evidence for directionally structured spiking, clustered bursting, partial coupling between spatial sectors, and slow propagation of electrical activity across the mycelial network.

\section{Materials and Methods}

Experiments were conducted using wood shavings substrate colonised by oyster fungi --- \textit{Pleurotus ostreatus} --- mycelium (Urban Farm-It, UK). The substrate was fully colonised and metabolically active at the time of recording. No external mechanical, chemical, thermal, or optical stimulation was applied during the experiments reported here. All recordings therefore
represent spontaneous electrical activity of the mycelial network under stable ambient conditions.

\begin{figure}[!tbp]
\centering
\includegraphics[width=0.75\linewidth]{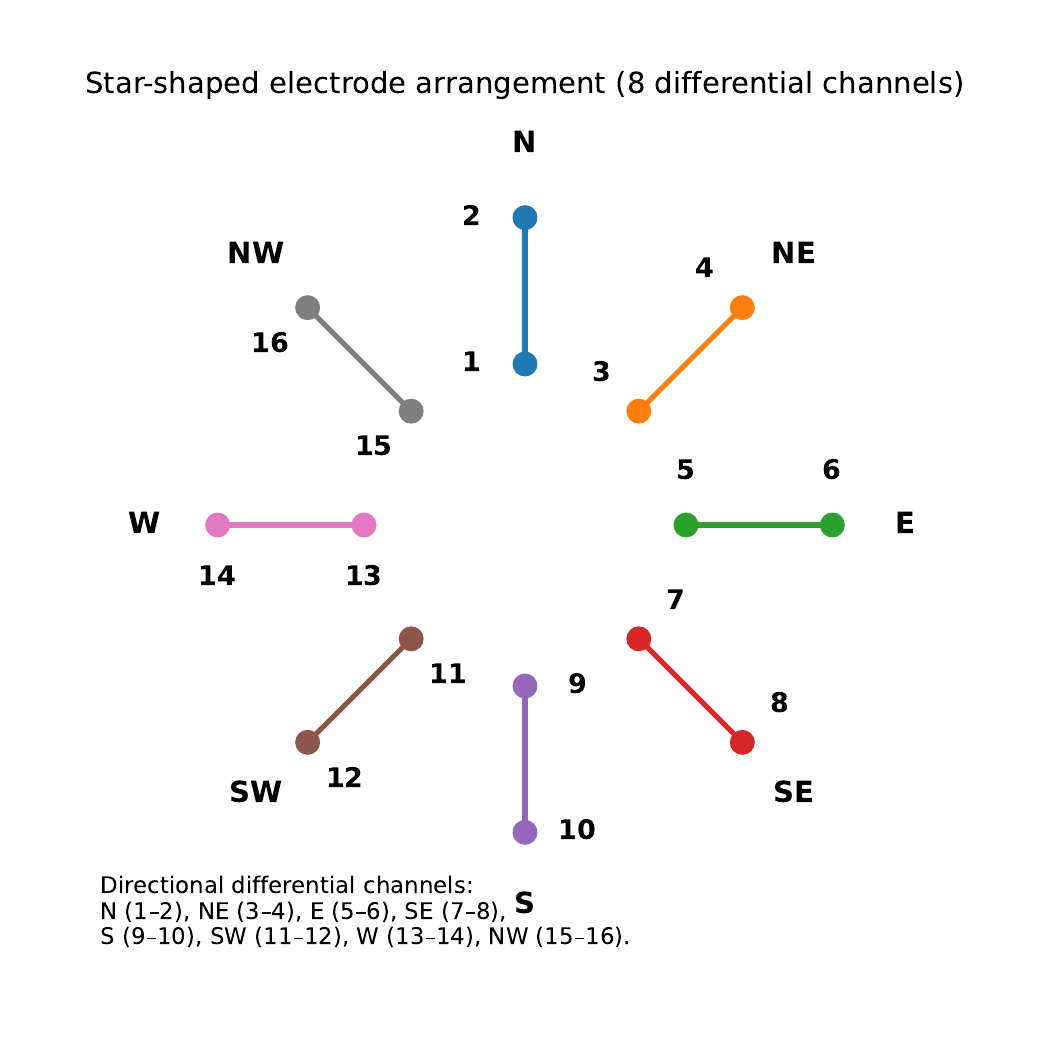}
\caption{Schematic of the electrode arrangement used for electrical recordings. Eight differential recording channels (1--8) were positioned at fixed locations within a substrate colonised by oyster fungi. The channels form an ordered spatial array, allowing comparison of spike onset times across neighbouring and non-neighbouring electrodes. This configuration enables detection of temporal delays and directional propagation of electrical activity along the mycelial network.}
\label{fig:electrode_schematic}
\end{figure}

Sixteen electrodes were inserted into the colonised substrate and arranged radially in a star-shaped geometry (Fig.~\ref{fig:electrode_schematic}).
The star-shaped electrode geometry was chosen to enable spatially resolved analysis of electrical activity without imposing a preferred linear direction on the system. Unlike linear or grid-based electrode arrangements, which implicitly bias measurements along one axis, the star geometry provides angular coverage around a central region of the colonised substrate. This configuration allows detection of directional asymmetries, sector-specific activity, and anisotropic propagation of electrical signals across the mycelial network. The star geometry also facilitates direct comparison between neighbouring and non-neighbouring channels at comparable radial distances from the centre, reducing confounding effects due to electrode spacing. By recording simultaneously from multiple directions, the arrangement enables discrimination between localised coupling, directional propagation, and global synchrony. This is particularly important for fungal systems, where growth and signal transmission may follow heterogeneous and dynamically evolving pathways within the substrate.

The electrodes were paired to form eight differential recording channels oriented along compass directions: North (1--2), North-East (3--4), East (5--6), South-East (7--8), South (9--10), South-West (11--12), West (13--14), and North-West (15--16). Differential recording was used to suppress common-mode noise and to emphasise local electrical potential
differences generated within the mycelium.

Electrical potential differences were recorded using a Pico ADC-24 high-resolution data logger (Pico Technology, St Neots, UK). The ADC-24 employs galvanically isolated differential inputs
and a 24-bit analogue-to-digital converter, providing high noise immunity and stable long-term recordings. The input impedance of the differential channels was approximately 2~M$\Omega$, with an offset error of tens of microvolts in the measurement range used. Signals were sampled at a rate of one sample per second. Internally, the data logger performs multiple measurements per second and stores their averaged value, reducing high-frequency noise while preserving
slow physiological dynamics.

Recordings were conducted continuously over extended periods up to five days, and three independent recording sessions were analysed. Raw data were exported from the Pico software environment and processed offline.

Data analysis was performed using custom scripts written in Python. Numerical processing relied on standard scientific libraries including NumPy and SciPy, while data handling and visualisation employed pandas and Matplotlib. All analyses were performed on detrended signals to remove slow baseline drift while preserving transient electrical events.

Electrical spikes were detected using adaptive thresholding based on signal dispersion within each channel. A spike was defined as a contiguous excursion above a channel-specific threshold exceeding baseline by a fixed multiple of signal dispersion and persisting for at least tens of seconds. Detected spikes were characterised by their timing and peak amplitude. Spike trains were analysed to compute spike counts, inter-spike intervals, and amplitude
distributions. Bursting behaviour was identified by grouping spikes into clusters separated by quiescent intervals exceeding a fixed temporal gap, allowing identification of trains of electrical activity.

Inter-channel coupling was quantified using Pearson correlation coefficients computed on normalised, detrended signals. Correlation matrices were used to assess the spatial structure of electrical coordination across directional channels. To analyse propagation of electrical activity, burst onsets were identified in a selected reference channel, and delays to the first subsequent spike in other channels were measured within a predefined temporal window. Propagation delays were summarised using median values across multiple burst events and interpreted directly in physical time using the known sampling rate.

\section{Results}

\begin{figure}[!tbp]
\centering
\includegraphics[width=0.95\linewidth]{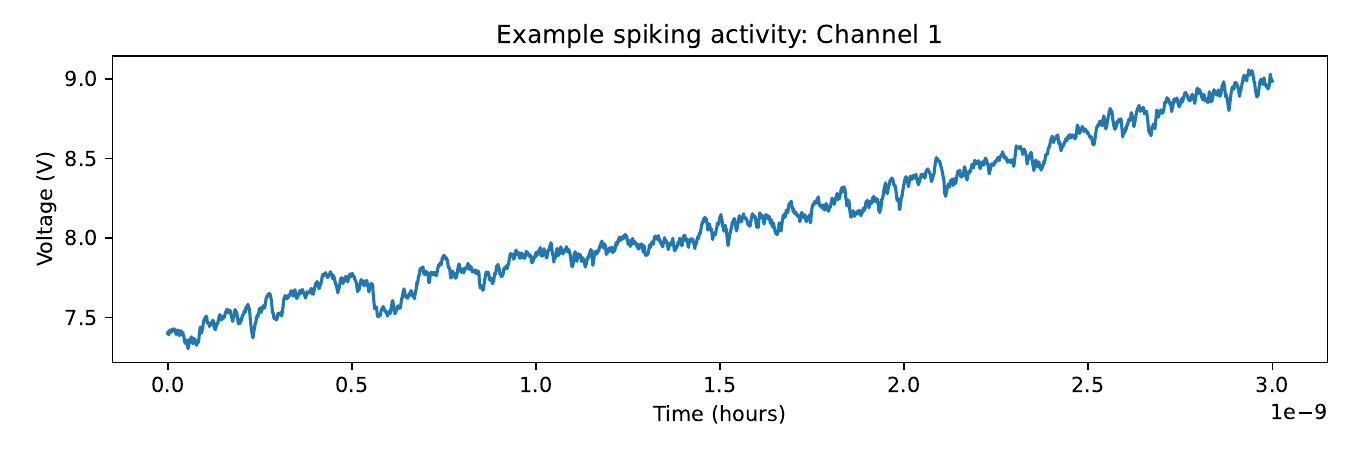}
\includegraphics[width=0.95\linewidth]{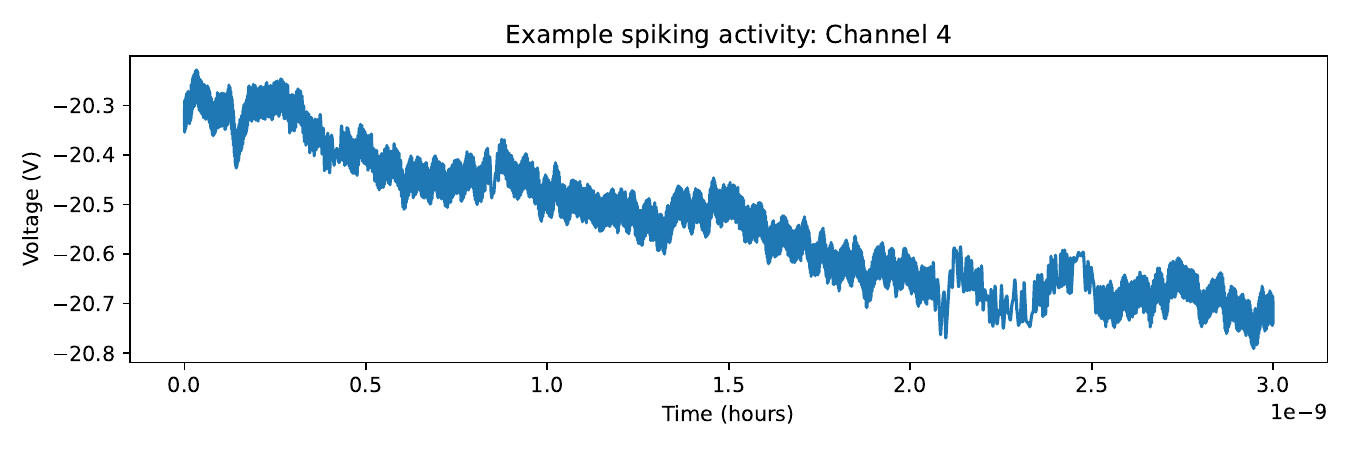}
\includegraphics[width=0.95\linewidth]{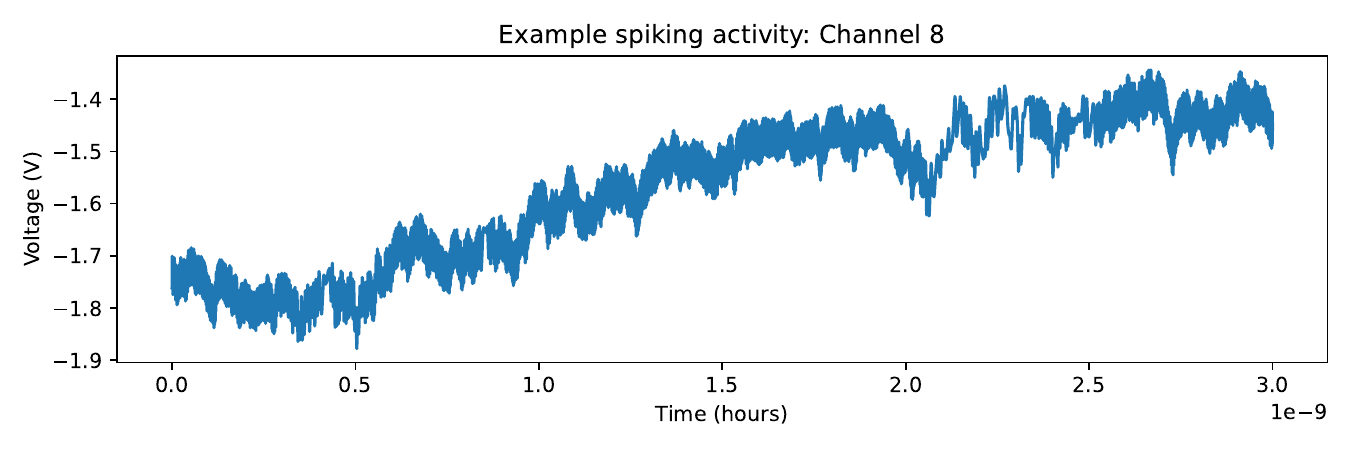}
\caption{Examples of electrical spiking activity recorded on different channels. Each panel shows raw electrical potential over a representative six-hour window for channels 1, 4, and 8, respectively. Spikes appear as slow, high-amplitude excursions occurring on time scales of minutes to hours. Differences in timing across channels illustrate spatiotemporal structure of the electrical activity.}
\label{fig:example_spikes}
\end{figure}

Electrical recordings from substrates colonised by oyster fungi revealed rich and diverse spiking activity across all channels. Representative examples of raw electrical signals recorded on different channels are shown in Fig.~\ref{fig:example_spikes}. Spikes appear as slow, high-amplitude excursions from baseline, occurring on time scales of minutes to hours, but their temporal patterns and morphologies vary substantially between channels. Some channels exhibit relatively isolated spikes separated by long silent periods, while others display clusters of closely spaced spikes or prolonged periods of elevated activity. Spike amplitudes, durations, and waveform shapes also differ across channels, indicating heterogeneous local dynamics within the colonised substrate. This diversity of spiking behaviour motivates the quantitative analyses presented below, which characterise individual spikes, spike trains, and the spatiotemporal relationships between channels.

\subsection{Directional Heterogeneity of Spiking Activity}

\begin{figure}[!tbp]
\centering
\includegraphics[width=\linewidth]{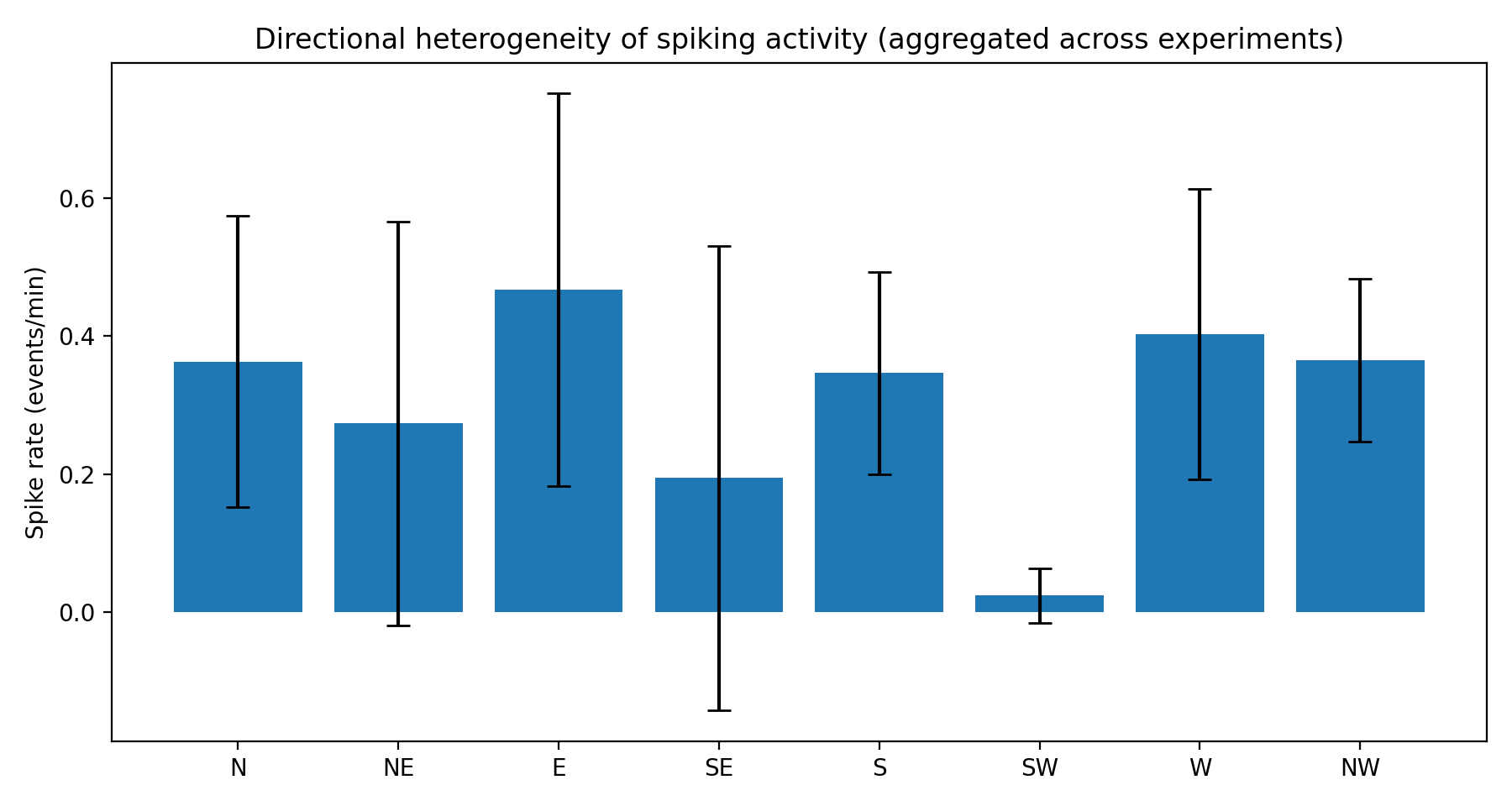}
\caption{Mean spike rate (events/min) by directional channel, aggregated across all recording
sessions. Error bars denote standard deviation between experiments. Electrical spiking is
strongly anisotropic, with order-of-magnitude differences between directions and substantial
inter-session variability.}
\label{fig:agg_spike_rates}
\end{figure}

Electrical spiking activity exhibited pronounced directional heterogeneity across the mycelial
network. When statistics were aggregated over all recording sessions, spike rates differed by
more than an order of magnitude between directional channels. Certain directions consistently
displayed high spiking activity, while others were weakly active or nearly silent, indicating
strong spatial anisotropy of electrical excitability.

Figure~\ref{fig:agg_spike_rates} shows the mean spike rate (events per minute) for each
directional channel, averaged across all experiments, with error bars denoting inter-session
variability. The distribution reveals a clear separation between highly active directions and
low-activity sectors. Notably, several directions exhibit consistently elevated spiking rates,
whereas others contribute minimally to the overall electrical activity. The relatively large
standard deviations reflect that the identity of the most active directions can vary between
recording sessions, suggesting dynamic reconfiguration of electrically active regions rather
than fixed spatial patterns.

\begin{figure}[!tbp]
\centering
\includegraphics[width=0.8\linewidth]{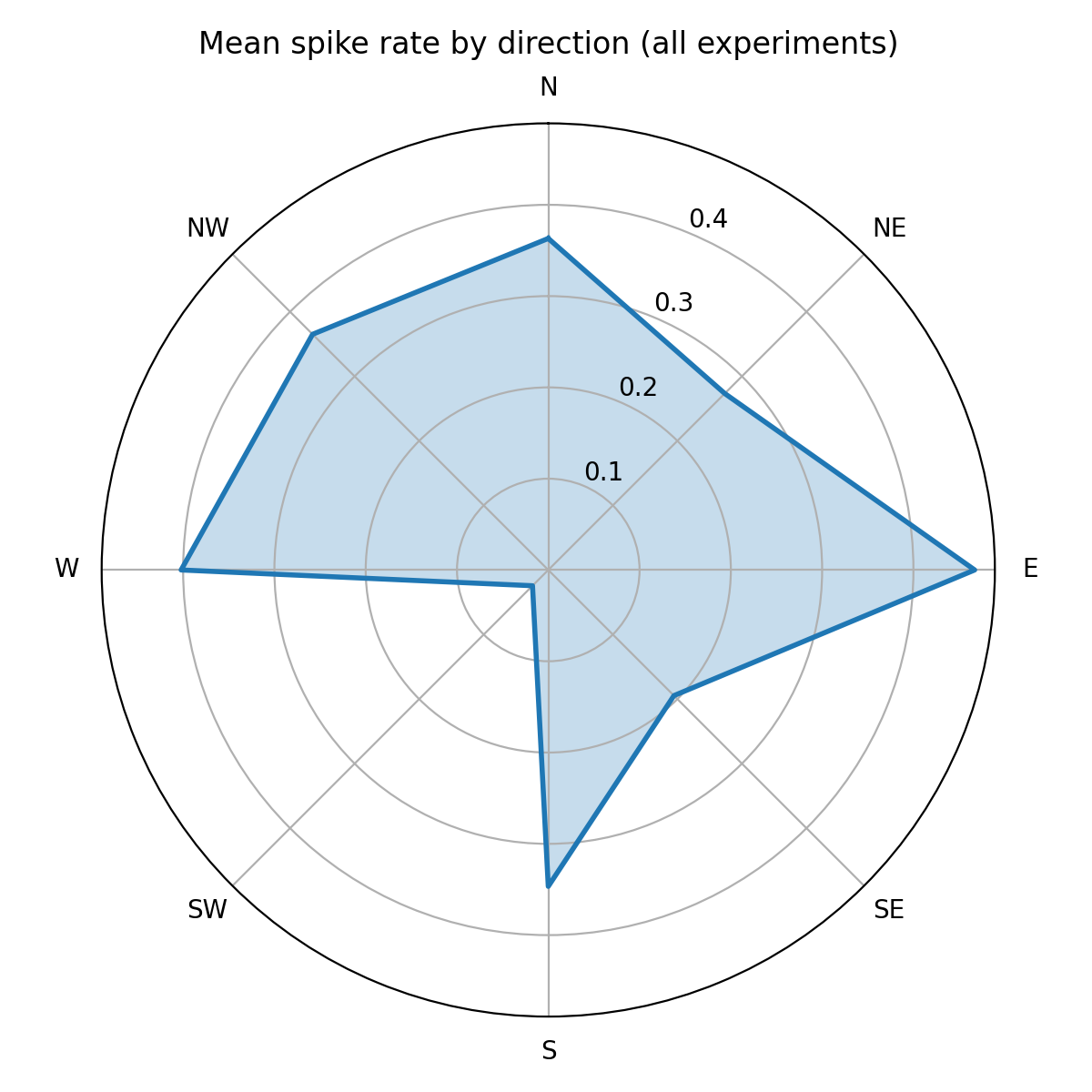}
\caption{Polar representation of mean spike rate by direction, aggregated across all
experiments. The star-electrode arrangement reveals pronounced angular anisotropy of electrical
activity within the mycelial network.}
\label{fig:agg_polar}
\end{figure}

The same data are represented in polar form in Fig.~\ref{fig:agg_polar}, which highlights the
angular structure of spiking activity. Rather than forming a uniform circular distribution, the
spike rates exhibit distinct lobes corresponding to preferred directions of electrical
activity. This polar representation emphasises that electrical signalling within the mycelium
is not isotropic but organised along spatially biased pathways.

Importantly, the observed heterogeneity persists despite pooling data from multiple independent
recordings, indicating that directional anisotropy is a robust feature of the system rather
than an artefact of a single experiment. At the same time, variability across sessions implies
that the spatial pattern of activity is not static but adapts over time, consistent with
reorganisation of physiological states, transport pathways, or local excitability within the
mycelial network.

Together, these results demonstrate that electrical spiking in oyster mycelium is spatially
structured and direction-dependent. The star-shaped electrode configuration enables detection
of this heterogeneity by resolving angular differences in activity, revealing a distributed
electrical system composed of dynamically reconfigurable active regions rather than uniform or
globally synchronised signalling.

\subsection{Spike Amplitudes and Bursting Dynamics}

\begin{figure}[!tbp]
\centering
\includegraphics[width=\linewidth]{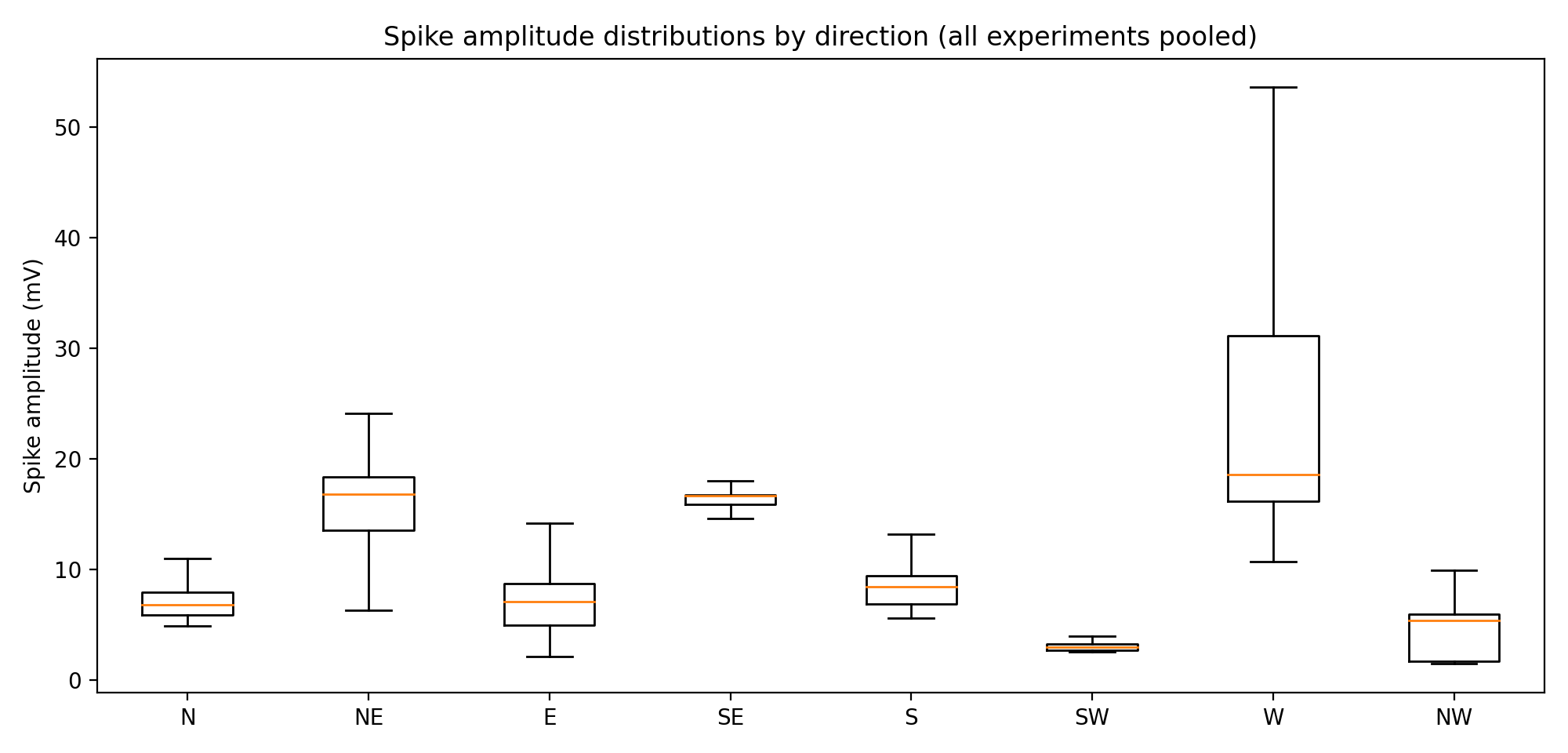}
\caption{Spike amplitude distributions by direction, pooled across all recording sessions.
Amplitude statistics differ substantially between directions, indicating that spatial sectors
of the mycelial network operate in distinct electrical regimes.}
\label{fig:amp_box}
\end{figure}

Electrical spike amplitudes varied over a wide range, spanning from a few millivolts to several
tens of millivolts, with rare events exceeding this range. When pooled across all recording
sessions, amplitude distributions differed substantially between directional channels, as
shown in Fig.~\ref{fig:amp_box}. Some directions exhibited relatively narrow amplitude
distributions dominated by low-amplitude spikes, whereas others showed broader distributions
with pronounced upper tails, indicating occasional high-amplitude events.

\begin{figure}[t]
\centering
\includegraphics[width=0.8\linewidth]{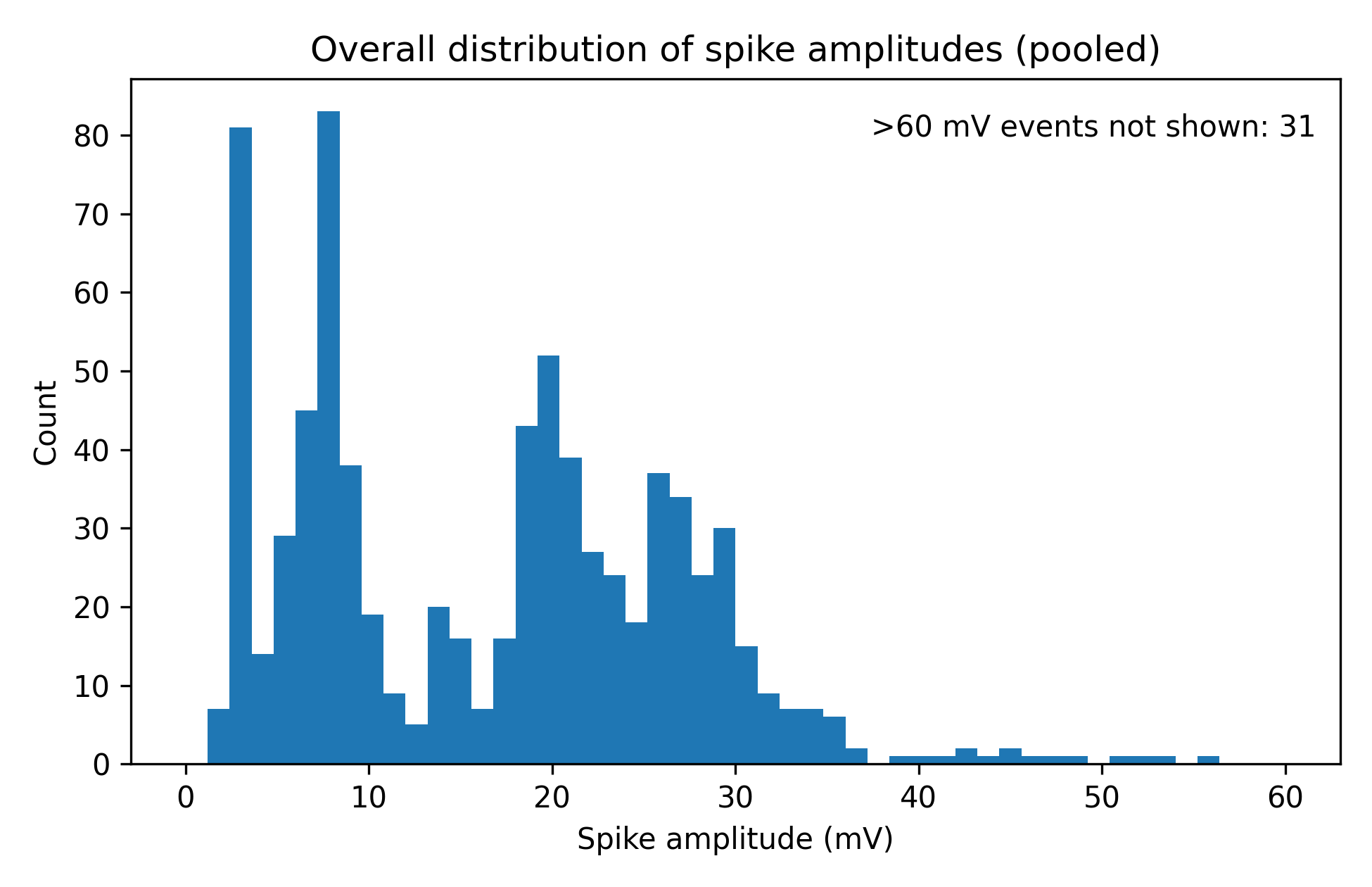}
\caption{Overall distribution of spike amplitudes pooled across all experiments. The distribution is heavy-tailed, with most events in the low-to-moderate millivolt range and
rare high-amplitude spikes reaching tens to over one hundred millivolts.}
\label{fig:amp_hist}
\end{figure}

The overall distribution of spike amplitudes (Fig.~\ref{fig:amp_hist}) is heavy-tailed, with
most events concentrated in the low-to-moderate millivolt range and a small number of large
events contributing disproportionately to the upper end of the distribution. Notably, some
directions with low overall spike frequency nevertheless produced high-amplitude spikes,
suggesting rare but intense episodes of electrical activity rather than sustained oscillatory
behaviour.

\begin{figure}[!tbp]
\centering
\includegraphics[width=0.85\linewidth]{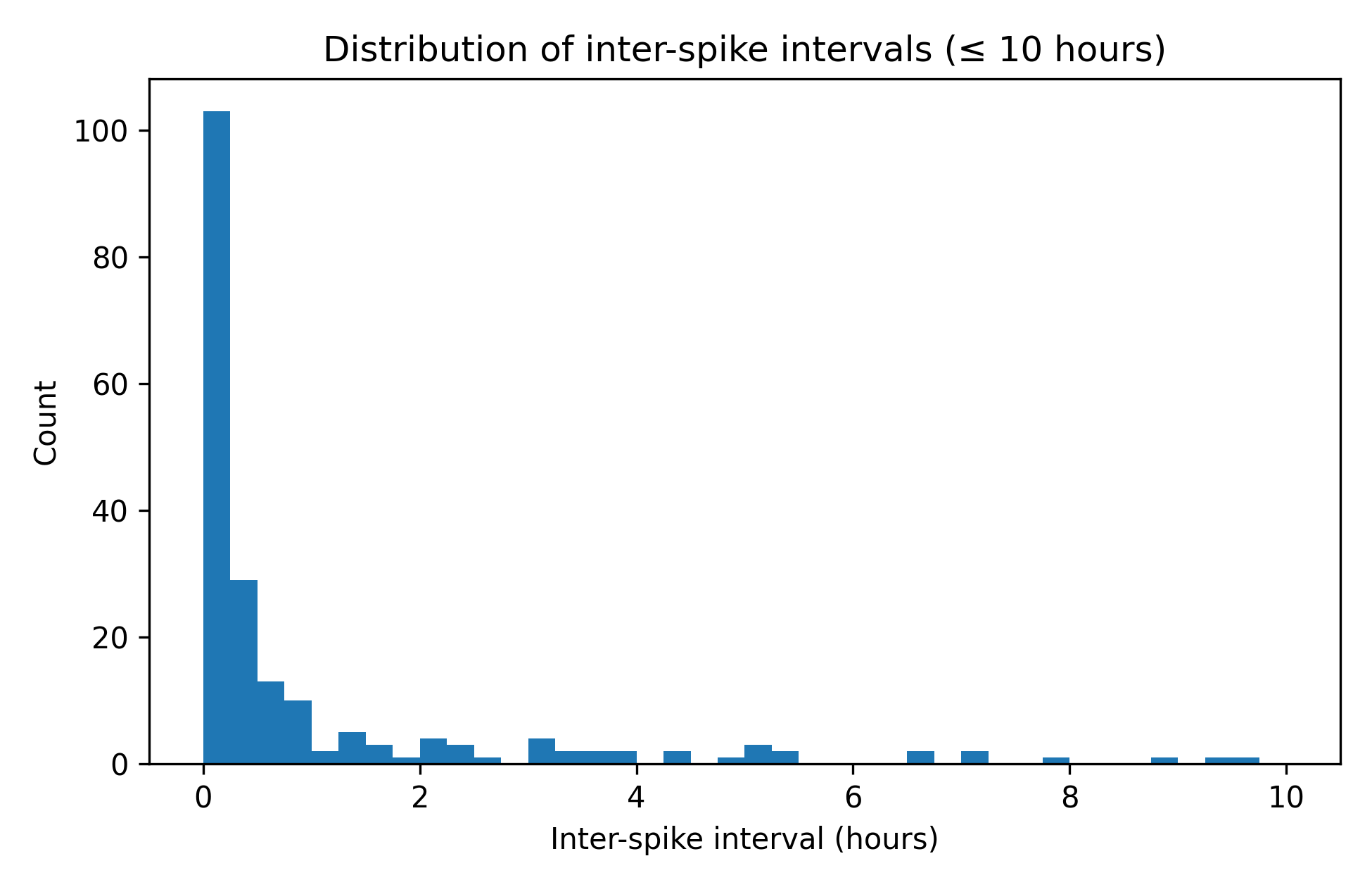}
\vspace{0.3cm}
\includegraphics[width=0.83\linewidth]{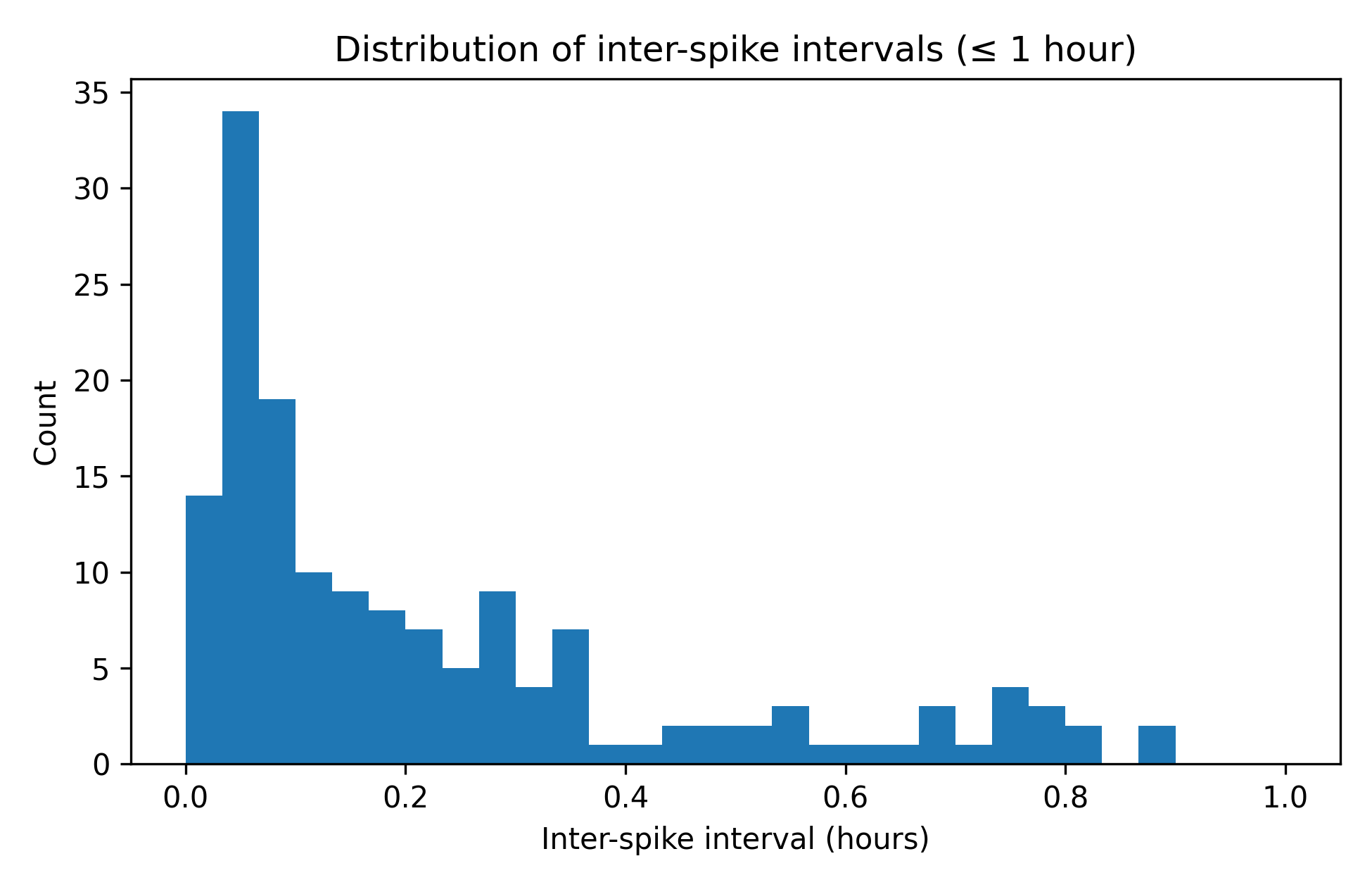}
\caption{Distributions of inter-spike intervals (ISIs) pooled across all experiments and channels, shown at multiple time scales. 
Top: ISI distribution restricted to intervals up to 10~hours, highlighting the structure of typical ISIs on sub-hour to few-hour time scales. 
Bottom: ISI distribution restricted to intervals up to 1~hour, resolving short-time-scale spike timing and showing that even the shortest ISIs occur on the order of minutes rather than seconds.}
\label{fig:isi_distributions}
\end{figure}

Spike timing was strongly non-uniform. Inter-spike interval analysis (Fig.~\ref{fig:isi_distributions})
reveals a concentration of short intervals alongside much longer gaps, inconsistent with
independent random spiking. Instead, spikes cluster in time, forming bursts separated by
quiescent periods lasting several minutes. This temporal organisation was observed across
multiple channels and recording sessions.

\begin{figure}[!tbp]
\centering
\includegraphics[width=\linewidth]{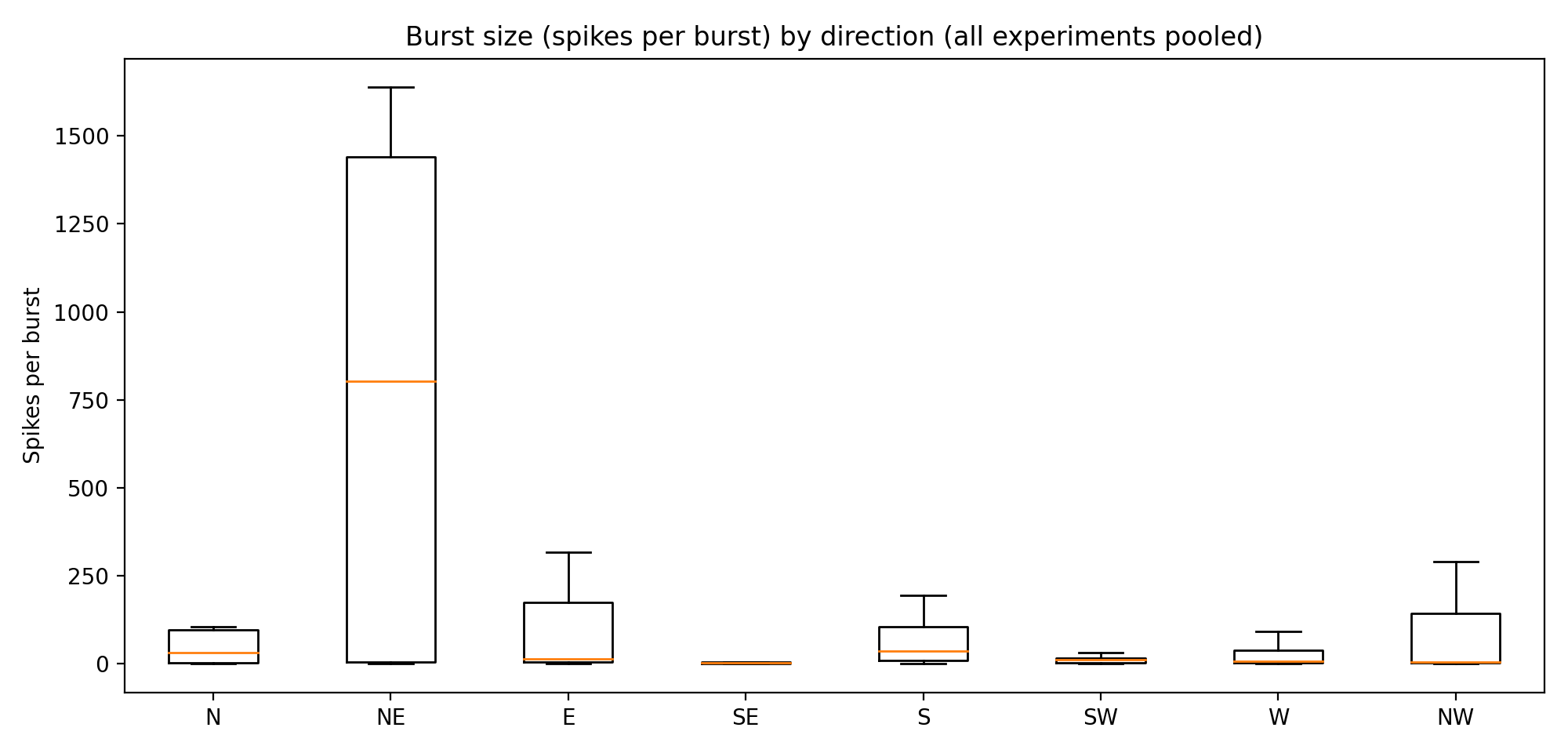}
\caption{Burst size (number of spikes per burst) by direction, pooled across all experiments.
Spikes cluster into trains separated by quiescent intervals, and burst size differs strongly
between spatial sectors.}
\label{fig:burst_size}
\end{figure}

\begin{figure}[t]
\centering
\includegraphics[width=\linewidth]{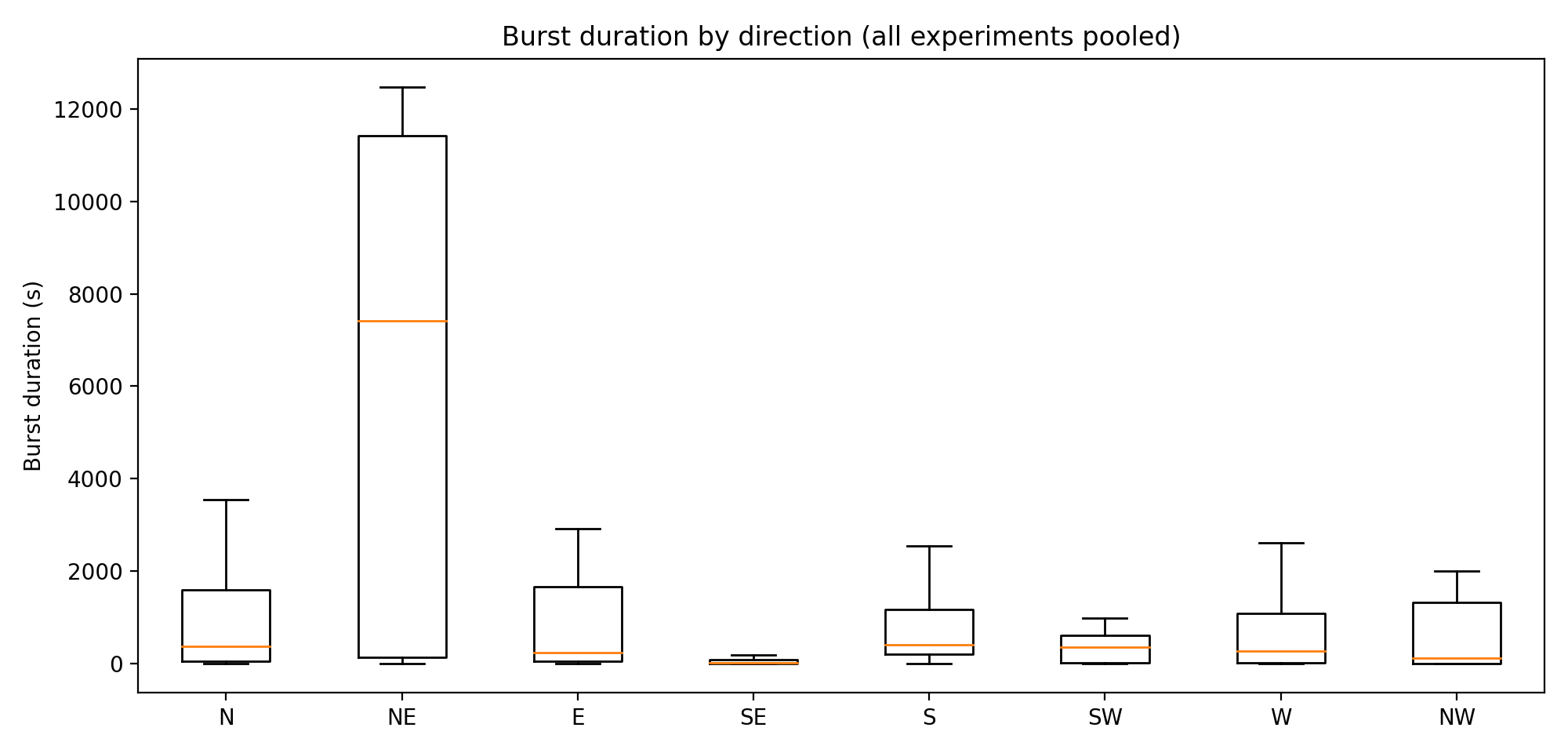}
\caption{Burst duration (seconds) by direction, pooled across all experiments.
Bursts range from short transient events to extended trains lasting many minutes, consistent
with excitable rather than purely stochastic dynamics.}
\label{fig:burst_dur}
\end{figure}

Bursting behaviour was quantified by grouping spikes into trains separated by extended
intervals. The number of spikes per burst varied widely and differed between directions
(Fig.~\ref{fig:burst_size}), indicating spatial heterogeneity in burst organisation. Burst
durations likewise spanned a broad range, from short transient events to extended trains
lasting many minutes (Fig.~\ref{fig:burst_dur}). Such variability is characteristic of
excitable systems operating far from steady-state equilibrium.

Taken together, the amplitude statistics and temporal clustering demonstrate that fungal
electrical activity is neither uniform nor stochastic. Instead, it consists of heterogeneous
spiking events organised into bursts whose size and duration depend on spatial location within
the mycelial network. These findings further support interpretation of the mycelium as a
distributed excitable medium capable of episodic, spatially structured electrical signalling.

\subsection{Inter-Channel Correlations}

To quantify coupling between recording sites, we computed pairwise correlation coefficients between electrical signals recorded on the eight channels. Correlation analysis revealed partial and spatially constrained coupling rather than global synchrony.

\begin{figure}[!tbp]
\centering
\includegraphics[width=0.8\linewidth]{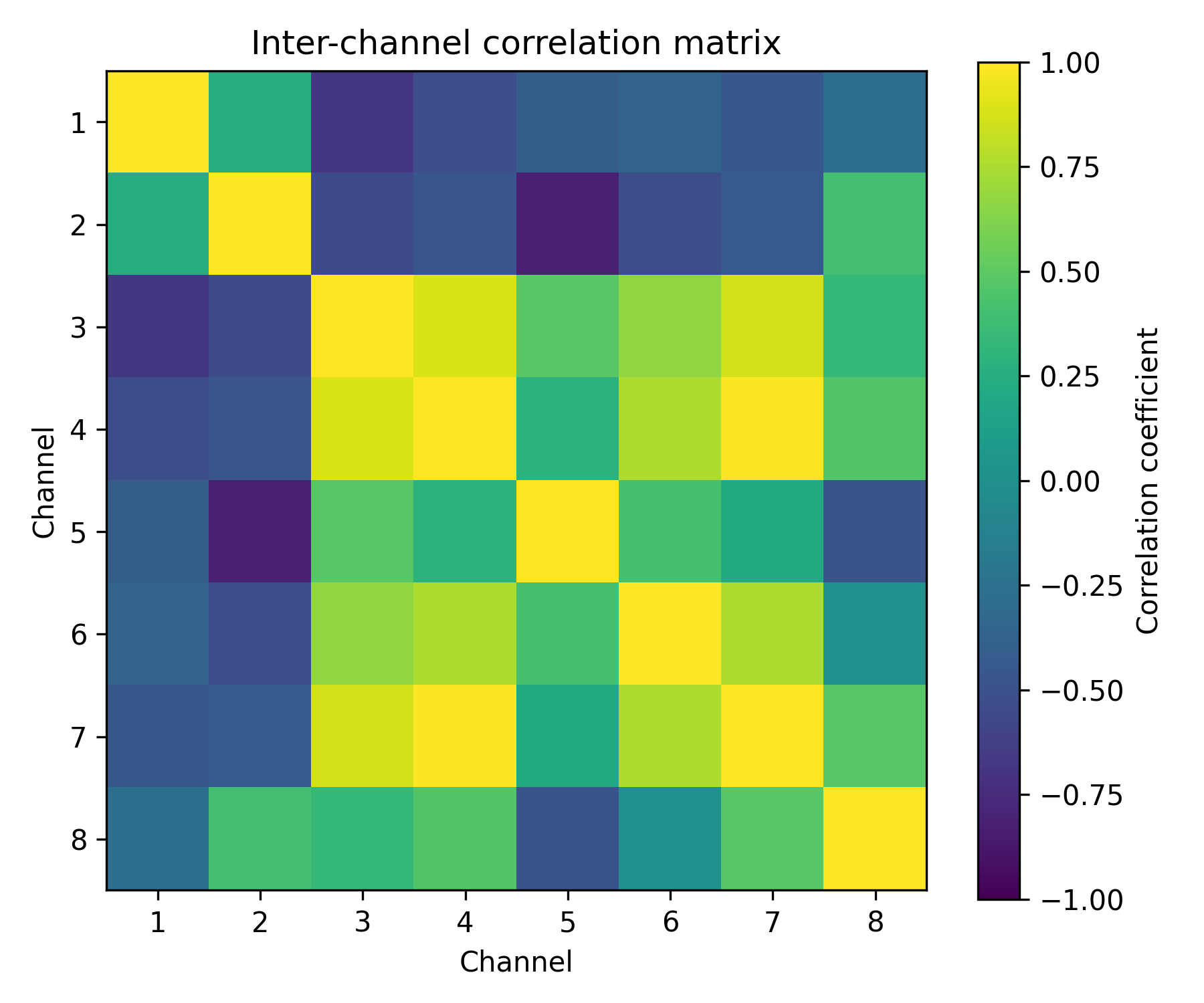}
\caption{Inter-channel correlation matrix for electrical activity recorded across the eight-channel array. Colour indicates pairwise correlation coefficients. Higher correlations are concentrated near the diagonal, corresponding to neighbouring channels, while distant channel pairs show weak or negligible correlations. The absence of uniformly high correlations indicates lack of global synchrony.}
\label{fig:correlation_matrix}
\end{figure}

The inter-channel correlation matrix is shown in Fig.~\ref{fig:correlation_matrix}. Higher correlation coefficients are concentrated near the diagonal of the matrix, corresponding to neighbouring channels in the ordered array. In contrast, channel pairs separated by larger distances exhibit weak or negligible correlations. No pattern of uniformly high correlations across all channels was observed, indicating the absence of system-wide oscillatory activity.

\begin{figure}[!tbp]
\centering
\includegraphics[width=0.7\linewidth]{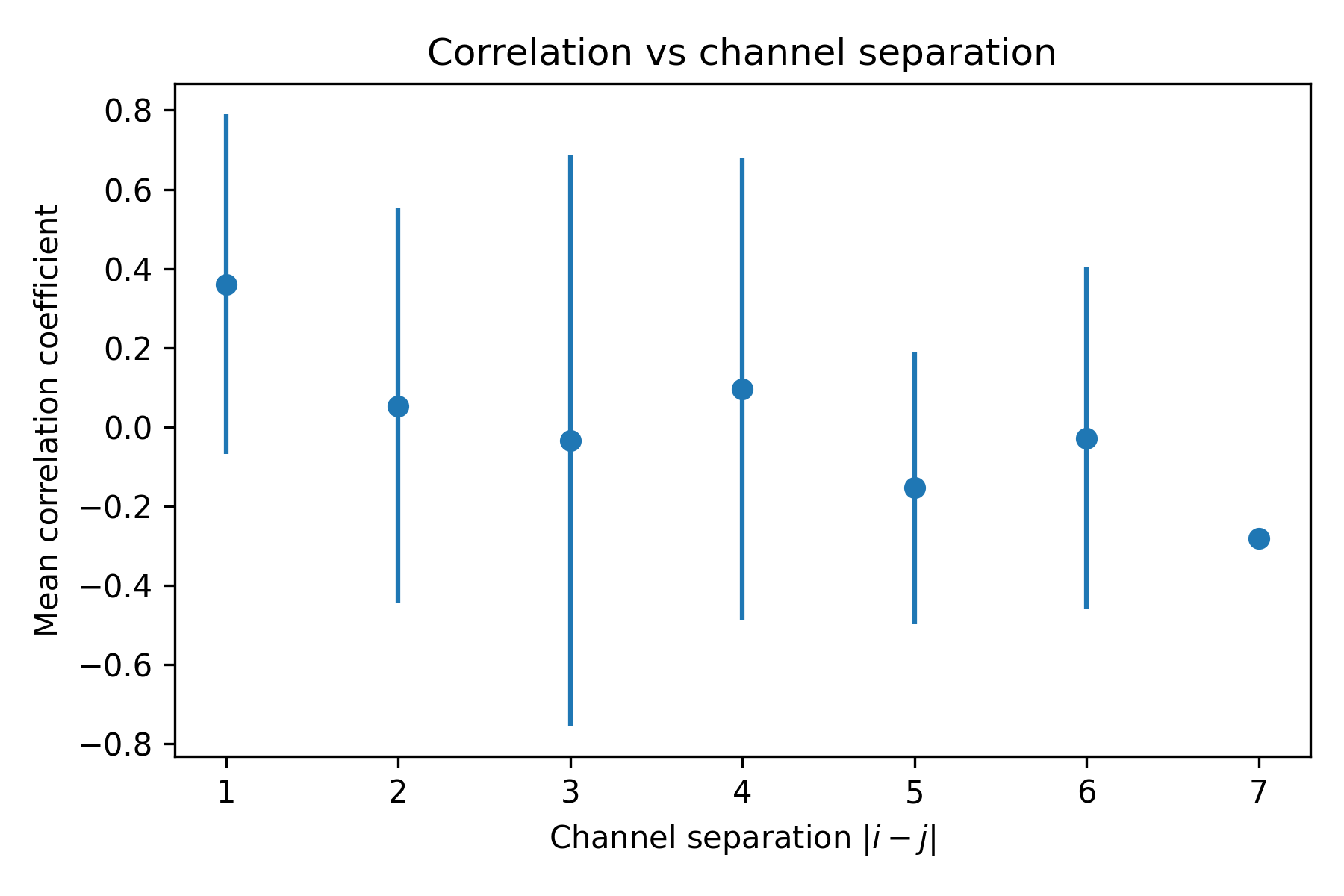}
\caption{Mean inter-channel correlation coefficient as a function of channel separation $|i-j|$. Correlation strength decreases with increasing distance between channels, indicating localised coupling of electrical activity constrained by spatial organisation. Error bars indicate standard deviation across channel pairs.}
\label{fig:correlation_distance}
\end{figure}

To examine the spatial dependence of coupling more directly, correlation coefficients were analysed as a function of channel separation $|i-j|$. As shown in Fig.~\ref{fig:correlation_distance}, the mean correlation coefficient decreases systematically with increasing channel separation. Neighbouring channels show the strongest coupling, while distant channel pairs exhibit markedly reduced correlations. This decay with distance is inconsistent with common-mode noise or global fluctuations, which would produce correlations largely independent of spatial separation.

Together, these results demonstrate that electrical activity in substrates colonised by oyster fungi is locally coordinated and constrained by spatial organisation along the electrode array. The observed correlation structure supports the interpretation that electrical signals propagate through the mycelial network and interact primarily with nearby regions rather than synchronising the system as a whole.

\subsection{Propagation of Electrical Bursts}

Event-based propagation analysis revealed that electrical bursts originating in one spatial
sector were followed by spiking activity in other directions with characteristic delays.
Propagation was quantified by measuring the time difference between the onset of burst
activity in a reference channel and the first subsequent spike detected in other channels,
using the known sampling rate of 1~Hz.

\begin{figure}[!tbp]
\centering
\includegraphics[width=\linewidth]{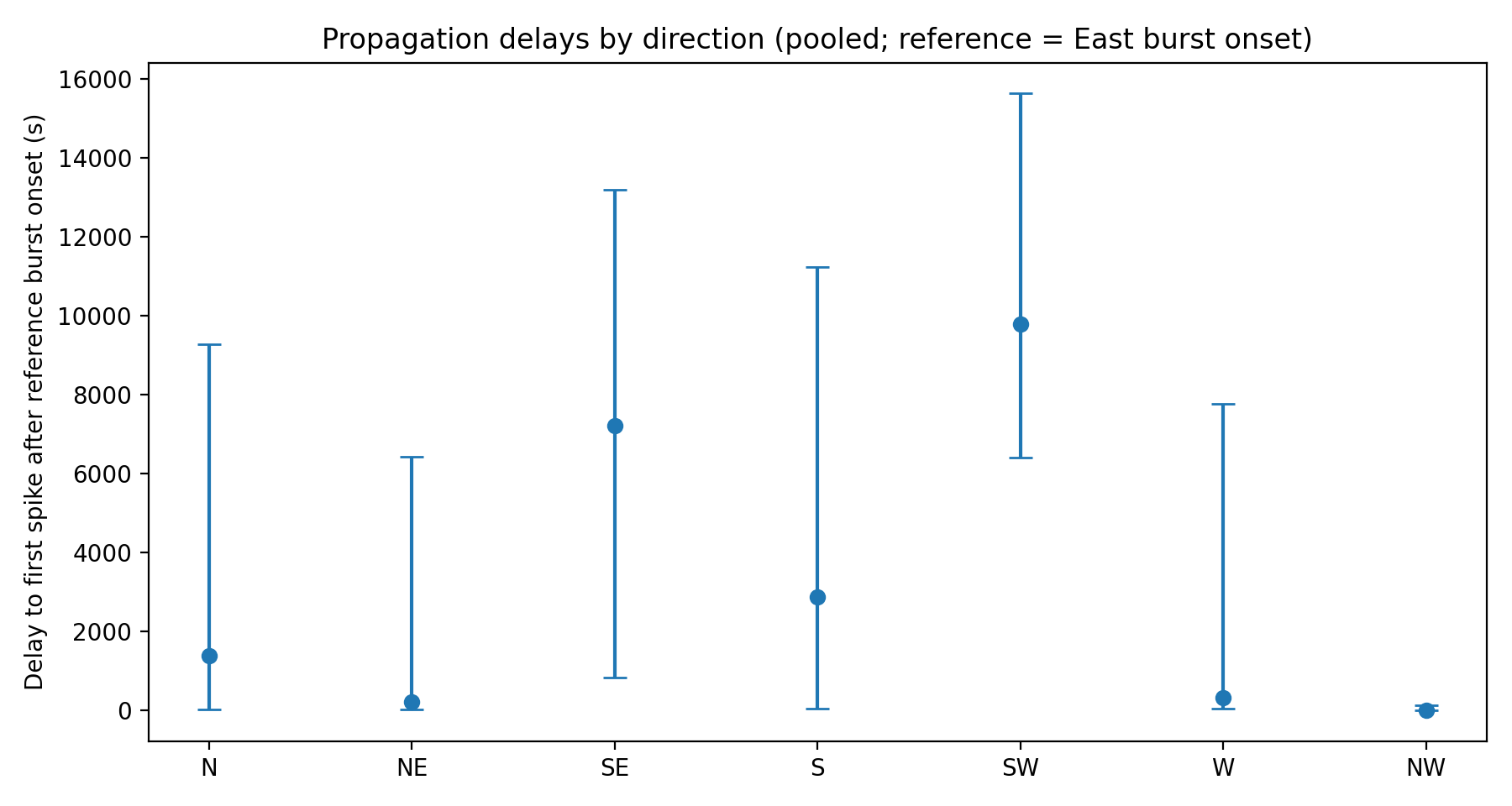}
\caption{Event-based propagation delays pooled across all experiments. Points show median delay
(seconds) from burst onset in the reference channel (East) to the first subsequent spike in
each direction; error bars show interquartile range. Recruitment occurs on timescales from
seconds to thousands of seconds, indicating slow physiological propagation.}
\label{fig:prop_delay_iqr}
\end{figure}

When pooled across all recording sessions, median propagation delays spanned several orders of
magnitude, ranging from a few seconds to several thousand seconds (Fig.~\ref{fig:prop_delay_iqr}).
In some directions, near-synchronous recruitment occurred within seconds of burst onset,
consistent with strong local coupling or shared physiological state. In contrast, other
directions exhibited delayed responses on the scale of tens of minutes to hours, indicating
weak coupling or indirect pathways of electrical recruitment.

\begin{figure}[!tbp]
\centering
\includegraphics[width=\linewidth]{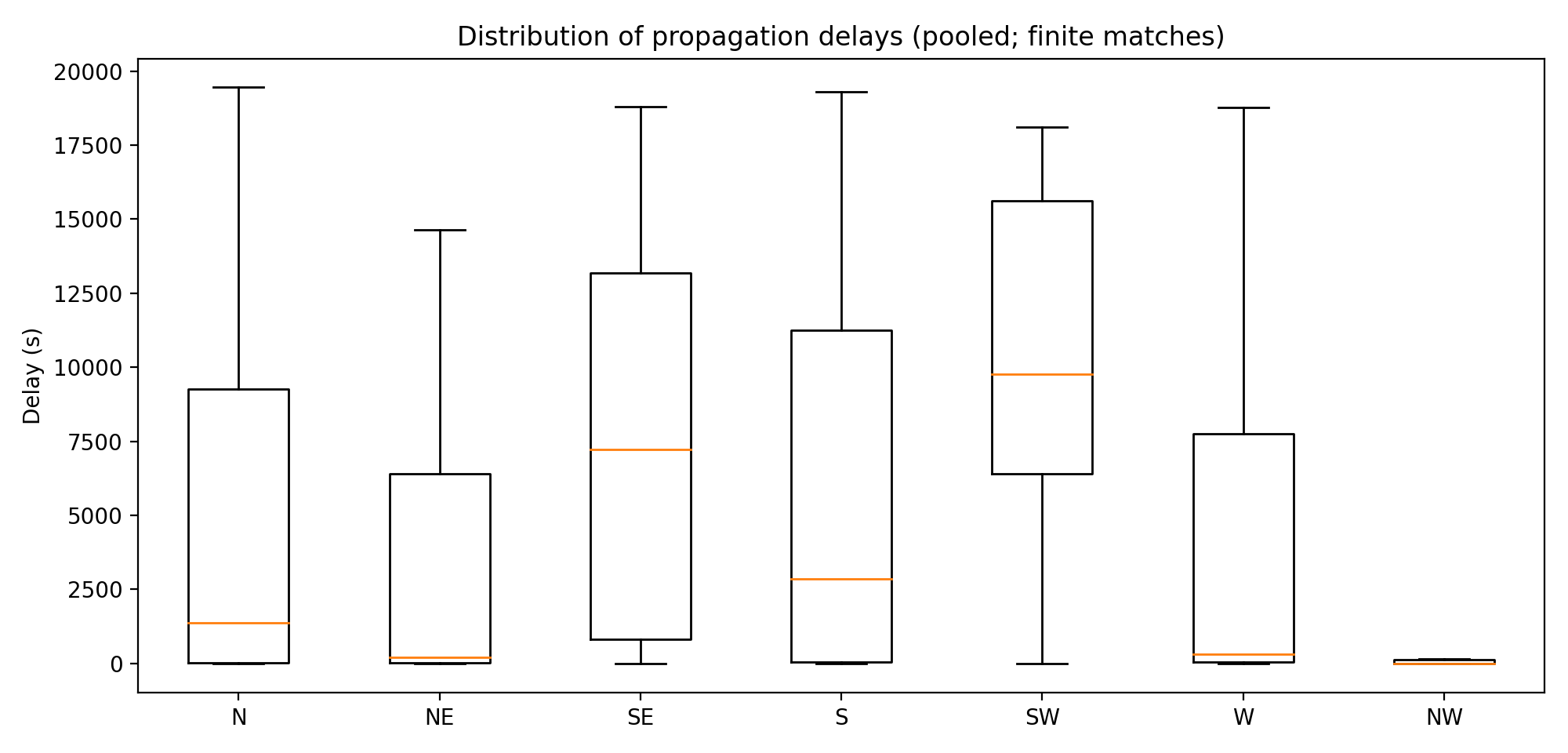}
\caption{Distribution of propagation delays (finite matched events only), pooled across all
experiments. Delay distributions are broad and heavy-tailed, reflecting variability in
recruitment timing characteristic of excitable media.}
\label{fig:prop_box}
\end{figure}

The full distributions of propagation delays (Fig.~\ref{fig:prop_box}) were broad and
heavy-tailed, with substantial variability between events. Such distributions are inconsistent
with rapid neuronal-like conduction along fixed pathways and instead suggest slow, state-dependent
processes governing signal spread through the mycelial network. These timescales are compatible
with physiological propagation mediated by growth dynamics, ionic fluxes, or metabolic
reconfiguration rather than fast excitable-wave transmission.

\begin{figure}[t]
\centering
\includegraphics[width=0.8\linewidth]{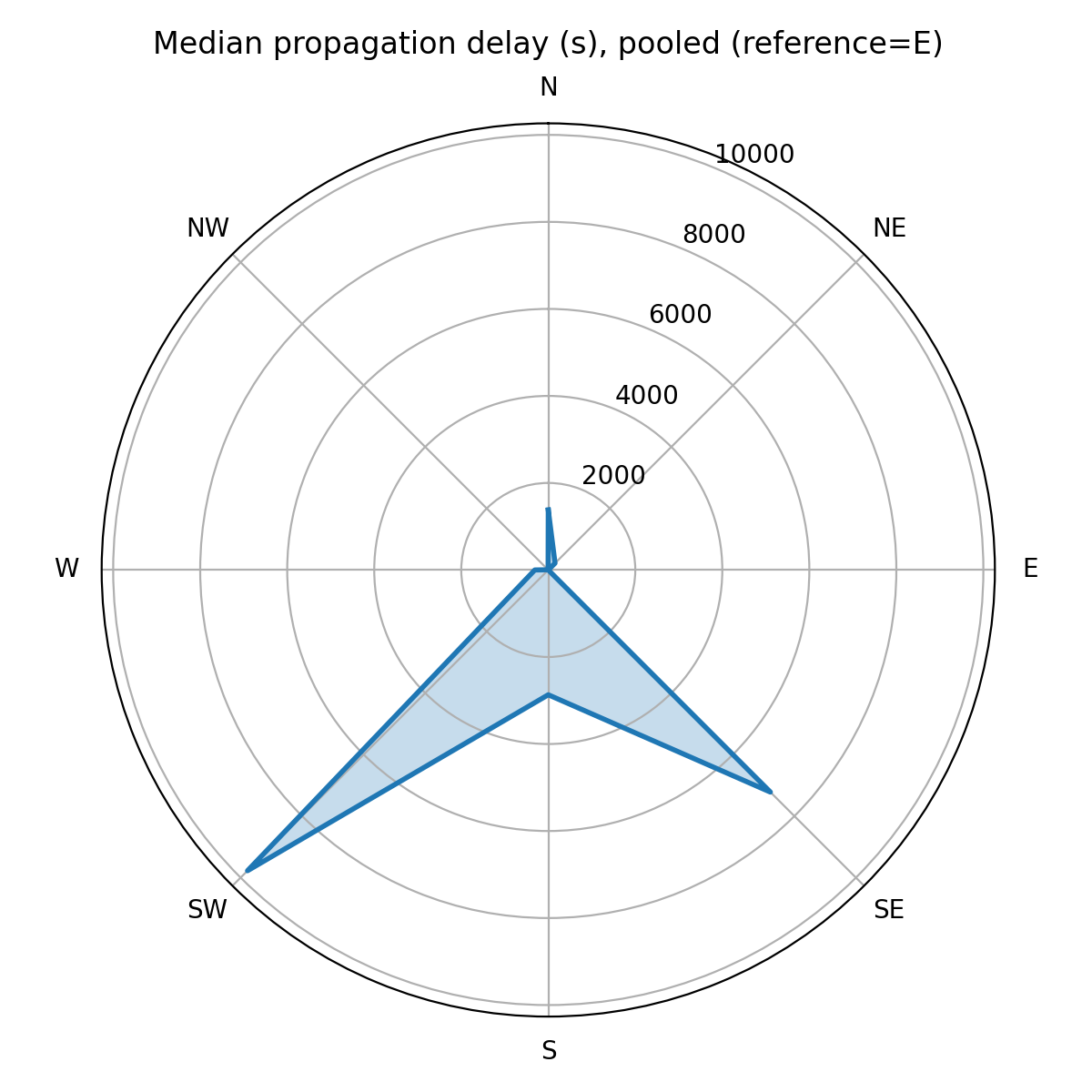}
\caption{Polar propagation map showing median recruitment delays (seconds) by direction pooled
across all experiments, relative to burst onset in the East channel. The non-uniform angular
structure indicates directional pathways of electrical recruitment.}
\label{fig:prop_polar}
\end{figure}

Directional structure in propagation is highlighted in the polar representation of median
delays (Fig.~\ref{fig:prop_polar}). Recruitment timing is strongly anisotropic, with some
directions responding rapidly and others exhibiting prolonged delays. This angular dependence
indicates preferred pathways of electrical influence within the mycelium, shaped by network
architecture and physiological heterogeneity.

\begin{figure}[!tbp]
\centering
\includegraphics[width=0.85\linewidth]{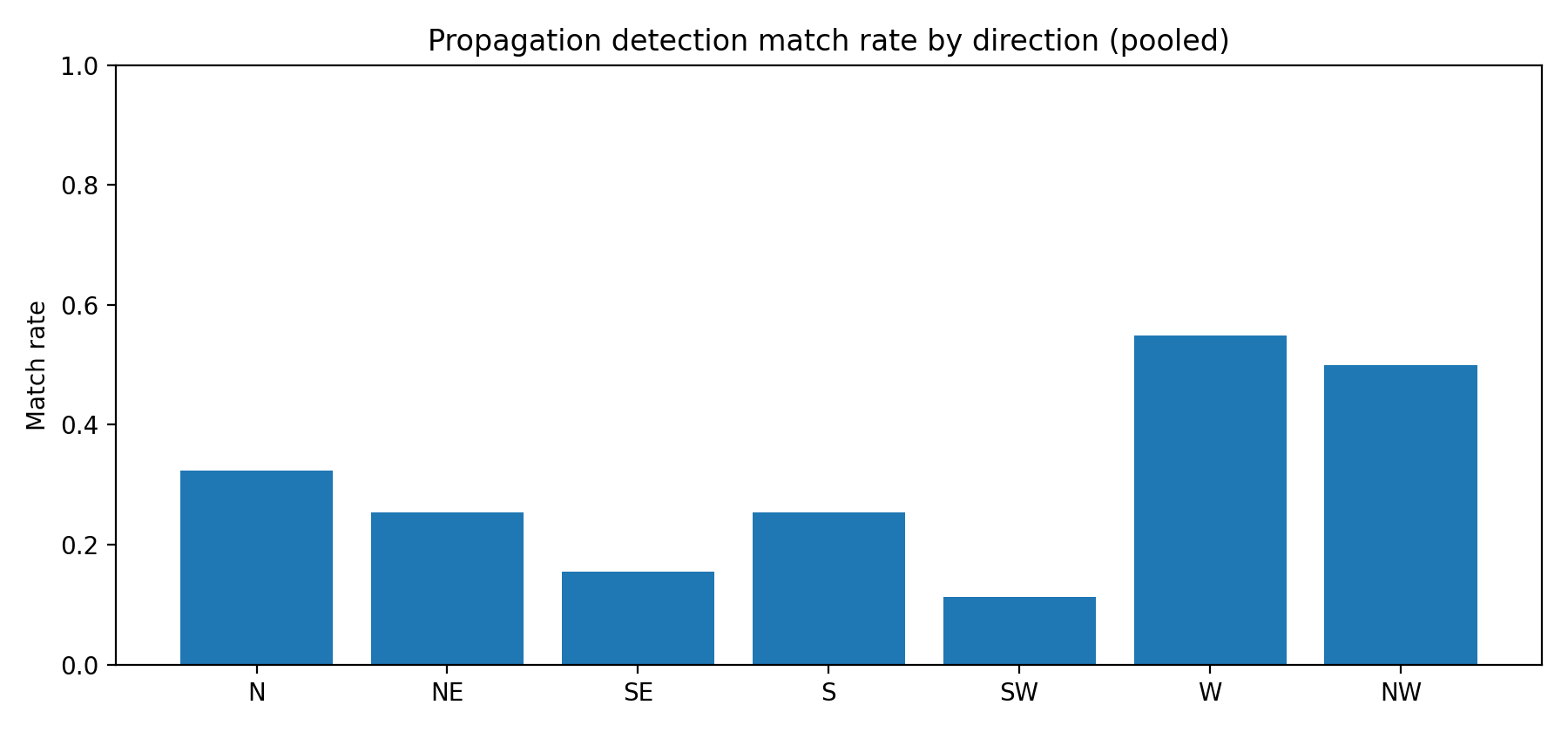}
\caption{Propagation detection match rate by direction pooled across all experiments. Match
rate is the fraction of reference burst onsets followed by at least one spike in the given
direction within the analysis window, providing an operational measure of coupling strength.}
\label{fig:prop_match}
\end{figure}

Propagation was not uniformly observed across all burst events. The fraction of reference
bursts followed by detectable activity in a given direction (match rate) varied
substantially between channels (Fig.~\ref{fig:prop_match}). Directions with low match rates
likely correspond to weakly coupled or sparsely active regions of the network, whereas higher
match rates indicate more reliable recruitment.

Although propagation patterns were reproducible within individual recording sessions, they
differed when data were pooled across sessions, reflecting changes in dominant pathways over
time. Together, these results demonstrate that electrical activity propagates through the
mycelial network via slow, spatially structured, and state-dependent processes, consistent with
distributed physiological signalling rather than rapid neural-like transmission.

\section{Discussion}

The results demonstrate that electrical activity in oyster mycelium is not spatially homogeneous
nor temporally random, but instead exhibits structured spiking, bursting, partial coupling, and
directed propagation. The absence of global synchrony, together with structured correlations and
delayed recruitment, argues against artefactual sources such as environmental noise or electrode
instability.

Directional heterogeneity of spiking activity suggests that different regions of the mycelium
operate in distinct physiological states. Such anisotropy likely reflects underlying variation
in hyphal density, nutrient availability, transport pathways, or metabolic demand. Importantly,
these spatial asymmetries persist over extended periods, indicating stable functional
organisation rather than transient fluctuations.

Bursting dynamics further support interpretation of the mycelium as an excitable medium.
Excitable systems naturally exhibit clustered responses separated by refractory periods.
In fungal networks, such bursts may correspond to coordinated transport events, metabolic
switching, or internally generated signalling. The presence of rare, high-amplitude spikes in
otherwise weakly active channels suggests episodic strong recruitment rather than continuous
oscillation.

Correlation analysis shows that electrical coupling is predominantly local. Adjacent spatial
sectors are more strongly correlated than distant ones, consistent with signalling constrained
by physical connectivity of hyphal cords and local ionic environments. Event-based propagation
analysis complements this picture by revealing slow recruitment of distant regions over
timescales of minutes to tens of minutes.

The star-shaped electrode geometry was essential in revealing these spatial features. Unlike
linear or sparse electrode arrangements, angular resolution allows detection of directional
biases, local coupling, and non-uniform propagation. Such spatially resolved approaches are
critical for advancing from descriptive observations of fungal electrical activity toward
mechanistic understanding.

While the present study does not attribute neural or cognitive function to fungal electrical
signalling, the observed dynamics demonstrate properties relevant to distributed information
processing, including excitability, signal propagation, temporal integration, and adaptive
reconfiguration. These findings support ongoing interest in fungi as living substrates for
biohybrid sensing and unconventional computation, as well as for understanding coordination in
biological systems lacking centralised control.

\section{Conclusion}

Using a star-shaped differential electrode array and recordings at one sample per second, we
have shown that electrical activity in oyster mycelium is directionally structured, burst-based,
partially correlated, and capable of slow propagation across spatial sectors. Electrical bursts
recruit different regions over timescales with characteristic delays ranging from seconds to minutes to hours, supporting a
model of distributed, slow coordination rather than rapid centralised control. These results
position fungal mycelium as a spatially extended excitable system and provide a foundation for
future studies of fungal physiology, biohybrid sensing, and unconventional information
processing.

\section*{Acknowledgements}

This project has received funding from the European Union's Horizon 2020 research and innovation programme FET OPEN ``Challenging current thinking'' under grant agreement No 858132. 

\bibliographystyle{plain}

\bibliography{biblio}

\end{document}